\documentclass[bib]{statapress}

\usepackage[crop,newcenter,frame]{pagedims}

\usepackage{sj}

\usepackage{epsfig}

\usepackage{stata}

\usepackage{amsmath,amssymb}
\usepackage{enumitem}
\usepackage{shadow}

\sjsetissue{$vv$}{$ii$}{$mm$}{$yyyy$}

\usepackage{threeparttable}
\usepackage{tabularx}

\newcommand{\transpose}{\text{\scalebox{0.7}{$\intercal$}}}
\newcommand{\convl}{\xrightarrow{L}}

\DeclareMathOperator{\rank}{rank}

\begin{document}
\inserttype[st0001]{article}
\author{Chen, Fang and Huang}{%
  Qihui Chen\\The Chinese University of Hong Kong, Shenzhen\\Shenzhen, China\\qihuichen@cuhk.edu.cn
  \and
  Zheng Fang\\Texas A\&M University\\College Station, TX\\zfang@tamu.edu
  \and
  Xun Huang\\San Francisco State University\\San Francisco, CA\\xhuang10@mail.sfsu.edu
}
\title[An Improved Test of Matrix Rank]{Implementing an Improved Test of Matrix Rank in Stata}
\maketitle

\begin{abstract}
We develop a Stata command, \texttt{bootranktest}, for implementing the matrix rank test of \citet{ChenFang2016Rank} in linear instrumental variable regression models. Existing rank tests employ critical values that may be too small, and hence may not even be first order valid in the sense that they may fail to control the Type I error. By appealing to the bootstrap, they devise a test that overcomes the deficiency of existing tests. The command \texttt{bootranktest} implements the two-step version of their test, and also the analytic version if chosen. The command also accommodates data with temporal and cluster dependence.

\keywords{\inserttag, \texttt{bootranktest}, test of matrix rank, bootstrap}
\end{abstract}

\section{Introduction}

In this paper, we present an introduction to the improved test of matrix rank developed by \citet{ChenFang2016Rank}, and then describe the Stata command \texttt{bootranktest} for implementing the test. The central message in \citet{ChenFang2016Rank} is that existing rank tests may fail to control Type I error, even in simple examples---see p.1791 of the paper. This in particular includes the influential test of \citet{Kleibergen_Paap2006rank} and conceivably other analogous tests such as \citet{Robin_Smith2000rank}. The reason is that existing tests set up the hypotheses as $\mathrm H_0: \mathrm{rank}(\Pi_0)= r$ (the null) versus $\mathrm H_1: \mathrm{rank}(\Pi_0)> r$ (the alternative) where $\Pi_0$ is the unknown matrix in question and $r$ is a prespecified rank. When the actual rank $\mathrm{rank}(\Pi_0)$ of $\Pi_0$ is strictly less than $r$ (a scenario covered by neither $\mathrm H_0$ nor $\mathrm H_1$), the distributions of a test statistic under $\mathrm{rank}(\Pi_0)= r$ may not stochastically dominate those under $\mathrm{rank}(\Pi_0)< r$. In practical terms, this means that the critical values employed by existing tests (designed for $\mathrm H_0: \mathrm{rank}(\Pi_0)= r$) may be too small, and hence may reject too often than desired. This motivates \citet{ChenFang2016Rank} to design a test with (asymptotically) correct rejection rates for $\mathrm H_0: \mathrm{rank}(\Pi_0)\le r$  versus $\mathrm H_1: \mathrm{rank}(\Pi_0)> r$, which is accomplished by estimating the null distributions of the test statistic through bootstrap. It turns out that the problem is nonstandard in the sense that the limiting null distributions under $\mathrm{rank}(\Pi_0)< r$ are nonstandard, and in particular depend on, among other things, whether the actual rank of $\Pi_0$ is smaller than the hypothesized value $r$. On the other hand, if $\mathrm{rank}(\Pi_0)= r$, their result uncovers the usual $\chi^2$ or $\chi^2$-type limiting distributions.

The remainder of this paper is organized as follows. Section \ref{Sec: Framework} lays out the framework of \citet{ChenFang2016Rank}, while Section \ref{Sec: Implementation} presents the implementation details. Section \ref{Sec: Installation} shows how to install the package, and Section \ref{Sec: the command} describes the command \texttt{bootranktest} in details. Section \ref{Sec: Examples} illustrates the use of the command in an empirical example. Throughout, we let $\mathbf M^{m\times k}$ be the collection of all $m\times k$ matrices. For $A\in\mathbf M^{m\times k}$, we denote the transpose of $A$ by $A^\transpose$. Finally, $I_k$ denotes the identity matrix of size $k$.

\section{Framework}\label{Sec: Framework}

\citet{ChenFang2016Rank} deal with a general unknown matrix $\Pi_0\in\mathbf M^{m\times k}$. Without loss of generality, we may assume $m\ge k$, by taking the transpose of $\Pi_0$ if necessary. The hypotheses are formulated as: for a prespecified integer $r$,
\begin{align}\label{Eqn: hypotheses1}
\mathrm H_0: \mathrm{rank}(\Pi_0)\le r \qquad \text{vs.}\qquad \mathrm H_0: \mathrm{rank}(\Pi_0)>r.
\end{align}
In many cases, $r=k-1$ which amounts to testing whether $\Pi_0$ has full (column) rank. This paper focuses on the canonical linear instrumental variable (IV) regression setup where $\Pi_0$ is a matrix of population coefficients in the first stage regression of a vector of endogenous variables on a vector of exogenous variables.

As a first step, we recall the singular value decomposition (SVD), which underlies the test of \citet{ChenFang2016Rank} as well as those of \citet{Robin_Smith2000rank} and \citet{Kleibergen_Paap2006rank}. The SVD may be viewed as a generalization of the eigendecomposition (or spectral decomposition) that are applicable to only square matrices to possibly rectangular matrices. When applied to $\Pi_0$, we may write
\begin{align}\label{Eqn: SVD}
\Pi_0= P_0\Sigma_0Q_0^\transpose ,
\end{align}
where $P_0\in\mathbf M^{m\times m}$ and $Q_0\in\mathbf M^{k\times k}$ are orthornormal matrices (i.e., they satisfy $P_0P_0^\transpose=I_m$ and $Q_0Q_0^\transpose=I_k$), and $\Sigma_0\in\mathbf M^{m\times k}$ is a diagonal matrix with its diagonal entries (called singular values of $\Pi_0$) in descending order. The singular values of $\Pi_0$ equal the square roots of the eigenvalues of the positive semidefinite matrix $\Pi_0^\transpose\Pi_0$, and hence are nonnegative. Since both $P_0$ and $Q_0$ are invertible (square) matrices, the rank of $\Pi_0$ is precisely the rank of $\Sigma_0$. Thus, a test for \eqref{Eqn: hypotheses1} may be based on $\Sigma_0$. To this end, we let $r_0\equiv\mathrm{rank}(\Pi_0)$ (the true rank of $\Pi_0$) and note the following block form of \eqref{Eqn: SVD}:
\begin{align}
\Pi_0= [P_{0,1},P_{0,2}]\underbrace{\begin{bmatrix}
  \sigma_{0,1} & \cdots &      0          &      0 & \cdots & 0\\
  \vdots          & \ddots & \vdots          & \vdots & \ddots & \vdots\\
  0               & \cdots & \sigma_{0,r_0} & 0      & \cdots & 0\\
  0               & \cdots & 0               & \sigma_{0,r_0+1}      & \cdots & 0\\
  \vdots          & \ddots & \vdots          & \vdots & \ddots & \vdots\\
  0               & \cdots & 0               & 0      & \cdots & \sigma_{0,k}\\
  0               & \cdots & 0               & 0      & \cdots & 0\\
  \vdots          & \ddots & \vdots          & \vdots & \ddots & \vdots\\
  0               & \cdots & 0               & 0      & \cdots & 0\\
\end{bmatrix}}_{\Sigma_0}[Q_{0,1},Q_{0,2}]^\transpose ,
\end{align}
where $\sigma_{0,j}\equiv\sigma_j(\Pi_0)$ is the $j$th largest singular value of $\Pi_0$, $P_{0,1}$ (resp.\ $P_{0,2}$) consists of the first $r_0$ (resp.\ last $k-r_0$) columns of $P_0$ associated with the nonzero (resp.\ zero) singular values of $\Pi_0$, and $Q_{0,1}$ and $Q_{0,2}$ are similar submatrices of $Q_0$. Thus, $\mathrm H_0$ holds if and only if the $k-r$ smallest singular values of $\Pi_0$ are zero. That is, $\mathrm H_0$ is equivalent to $\sigma_{r+1}(\Pi_0)=\cdots=\sigma_k(\Pi_0)=0$. Therefore, \eqref{Eqn: hypotheses1} may be equivalently formulated as
\begin{align}\label{Eqn: hypothesis2}
\mathrm H_0: \phi(\Pi_0)=0 \qquad \text{vs.} \qquad  \mathrm H_1: \phi(\Pi_0)>0,
\end{align}
where $\phi(\Pi_0)\equiv\sum_{j=r+1}^k\sigma_j^{2}(\Pi_0)$ is simply the sum of the $k-r$ smallest squared singular values of $\Pi_0$. While there are other possible choices other than taking squares, the formulation in \eqref{Eqn: hypothesis2} leads to $\chi^2$-type limiting distributions when $\mathrm{rank}(\Pi_0)=r$, as derived in \citet{Robin_Smith2000rank} and \citet{Kleibergen_Paap2006rank}.

In light of \eqref{Eqn: hypothesis2}, \citet{ChenFang2016Rank} employ the recaled plug-in statistic $n \phi(\hat\Pi_n)$ for an estimator $\hat\Pi_n$ of $\Pi_0$, and reject $\mathrm H_0$ if $ n \phi(\hat\Pi_n)$ is large (i.e., larger than the critical value). In the linear IV setup, $\hat\Pi_n$ is just the ordinary least square (OLS) estimator. To obtain the critical value, \citet{ChenFang2016Rank} establish that
\begin{align} \label{Eqn: asymp}
n\phi(\hat{\Pi}_{n})\overset{L}{\rightarrow} \sum_{j=r-r_{0}+1}^{k-r_{0}}\sigma^{2}_j(P_{0,2}^\transpose \mathcal{M} Q_{0,2}),
\end{align}
where $\convl$ signifies convergence in distribution, and $\mathcal M\in\mathcal M^{m\times k}$ is the asymptotic distribution of $\hat\Pi_n$, i.e., $\sqrt n\{\hat\Pi_n-\Pi_0\}\convl \mathcal M$. Thus, just like we use quantiles of the standard normal distribution as critical values for the standard $t$-test, we may set the critical values for $n \phi(\hat\Pi_n)$ to be the quantiles of the limiting distribution in \eqref{Eqn: asymp}, only that it is unknown due to the presence of the unknowns $\mathcal M$, $P_{0,2}$, $Q_{0,2}$ and critically $r_0$. The result \eqref{Eqn: asymp}, however, also suggests that this distribution may be estimated by replacing these unknowns with their estimated counterparts.

The estimation of (the distribution of) $\mathcal M$ may be accomplished by bootstrap. In the linear IV example, we may simply take the bootstrap estimator $\hat{\mathcal M}_n^*=\sqrt n\{\hat\Pi_n^*-\hat\Pi_n\}$, where $\hat\Pi_n^*$ is the first stage OLS estimator based on the bootstrap sample (rather than the original data). The matrices $P_{0,2}$ and $Q_{0,2}$ may be estimated by their sample analogs $\hat P_{n,2}$ and $\hat Q_{n,2}$ based on the SVD of $\hat\Pi_n$ for a given estimator $\hat r_n$ of $r_0$. Depending on how $\hat r_n$ is obtained, \citet{ChenFang2016Rank} develop two versions of their test. One estimator is constructed by counting the number of singular values of $\hat\Pi_n$ that are ``different'' from zero. For this, we need a threshold parameter $\kappa_n$ that is of smaller magnitude than $n^{-1/2}$ (e.g., $\kappa_n=n^{1-/4}$). Then we may set
\begin{align}\label{Eqn: rank estimator}
\hat r_n=\max\{j=1,\ldots,r: \sigma_j(\hat\Pi_n)\ge \kappa_n\}
\end{align}
if the set is nonempty and $\hat r_n=0$ otherwise. In turn, for a significance level $\alpha\in(0,1)$ (e.g., $\alpha=5\%$), one may set the critical value to the $1-\alpha$ conditional quantile (given the data), denoted by $\hat c_{n,1-\alpha}$, of
\begin{align}\label{Eqn: bootstrap}
\sum_{j=r-\hat r_n+1}^{k-\hat{r}_{n}}\sigma^{2}_j(\hat{P}_{2,n}^\transpose \hat{\mathcal M}_n^* \hat{Q}_{2,n}),
\end{align}
and reject $\mathrm H_0$ if $n\phi(\hat\Pi_n)>\hat c_{n,1-\alpha}$. This is called the analytic approach. Alternatively, one may obtain $\hat r_n$ via a sequential testing procedure based on the test of \citet{Kleibergen_Paap2006rank} at a user-specified level $\beta\in(0,\alpha)$. Such an estimator coincides with the true rank $r_0$ of $\Pi_0$ with probability $1-\beta$ (a high probability) in large samples. If $\hat r_n>r$, reject $\mathrm H_0$. Otherwise, take $1-\alpha+\beta$ conditional quantile $\hat c_{n,1-\alpha+\beta}$ of \eqref{Eqn: bootstrap}, and reject $\mathrm H_0$ if $n\phi(\hat\Pi_n)>\hat c_{n,1-\alpha+\beta}$. This is called the two-step approach. In both cases, the asymptotic null rejection rates are shown to be no larger than $\alpha$, regardless of whether $\mathrm{rank}(\Pi_0)=r$ or $\mathrm{rank}(\Pi_0)<r$.

\section{Implementation details}\label{Sec: Implementation}
As mentioned previously, we focus on the canonical linear IV regression setup. Specifically, we consider the following first stage regression
\begin{align}\label{Eqn: firststage}
X = \Pi_{0}^{\transpose}Z + \Gamma_{0}^{\transpose}W + u,
\end{align}
where $X\in\mathbf{R}^{k}$ is a vector of endogenous variables, $Z\in\mathbf{R}^{m}$ is a vector of nonconstant exogenous IVs, $W\in\mathbf{R}^{\ell}$ is a vector of exogenous controls which may contain a constant, and $u$ is the error term. Let $\{X_{i},Z_{i},W_{i}\}_{i=1}^{n}$ be a sample of $(X,Z,W)$, which may exhibit temporal or cluster dependence. The OLS estimator $\hat{\Pi}_n$ of $\Pi_0$ is
\begin{align}\label{Eqn: OLS}
\hat{\Pi}_n = (\mathbf{Z}^{\transpose}M_{\mathbf{W}}\mathbf{Z})^{-1}\mathbf{Z}^{\transpose}M_{\mathbf{W}}\mathbf{X},
\end{align}
where $M_{\mathbf{W}} = I_{n} - \mathbf{W}(\mathbf{W}^{\transpose}\mathbf{W})^{-1}\mathbf{W}$, $\mathbf{X}=(X_1,\ldots,X_{n})^{\transpose}$, $\mathbf{Z}=(Z_1,\ldots,Z_{n})^{\transpose}$, and $\mathbf{W}=(W_1,\ldots,W_{n})^{\transpose}$. To obtain the bootstrap estimator $\hat{\Pi}_n^{\ast}$, we use the residual based bootstrap. Let $\{\hat{u}_{i}\}_{i=1}^{n}$ be the residuals from the OLS estimation of \eqref{Eqn: firststage} and
\begin{align}\label{Eqn: bootfirststage}
X^{\ast}_{i} = \hat\Pi_{n}^{\transpose}Z_i + \hat\Gamma_{n}^{\transpose}W_i + \hat{u}^{\ast}_{i},
\end{align}
where $\hat\Gamma_{n}$ is the OLS estimator of $\Gamma_0$ and $\hat{u}^{\ast}_{i}$ is constructed as follows. In the case of independence data, we use the wild bootstrap \citep{Wu1986WildBoot} by setting $\hat{u}^{\ast}_{i} = \eta_i\cdot\hat{u}_{i}$, where $\{\eta_{i}\}_{i=1}^{n}$ are independent draws from $N(0,1)$. In the presence of cluster dependence, we use the cluster bootstrap \citep{CameronGelbachMiller2008BootCluster} and set $\hat{u}^{\ast}_{i} = \eta_g\cdot\hat{u}_{i}$ for $i$ in the $g$-th cluster, where $\{\eta_{g}\}_{g=1}^{G}$ are i.i.d. random variables with $P(\eta_g=-1)=P(\eta_g=1)=0.5$ and $G$ is the number of clusters. In the case of time series with temporal dependence, we use the block bootstrap \citep{Kunsch1989Jackknife}. Once a bootstrap sample $\{X^{\ast}_{i},Z_{i},W_{i}\}_{i=1}^{n}$ is obtained, the bootstrap estimator $\hat{\Pi}_n^{\ast}$ is given by
\begin{align}\label{Eqn: BootOLS}
\hat{\Pi}_n^{\ast} = (\mathbf{Z}^{\transpose}M_{\mathbf{W}}\mathbf{Z})^{-1}\mathbf{Z}^{\transpose}M_{\mathbf{W}}\mathbf{X}^{\ast},
\end{align}
where $\mathbf{X}^{\ast}=(X^{\ast}_1,\ldots,X^{\ast}_{n})^{\transpose}$.

In the following, we provide details for the implementation steps.

\noindent\underline{Step 1:} Compute the estimator $\hat\Pi_n$ in \eqref{Eqn: OLS} and the test statistic $n\phi(\hat\Pi_n)=\sum_{j=r+1}^k\sigma_j^{2}(\hat\Pi_n)$.

\noindent\underline{Step 2:} Construct the critical value for a given significance level $\alpha\in(0,1)$.
\begin{enumerate}[label=(\roman*)]
  \item Obtain $B$ bootstrap samples from the data (each of size $n$) according to \eqref{Eqn: bootfirststage}, compute $\hat\Pi_{n,b}^*$ according to \eqref{Eqn: BootOLS} based on the $b$th bootstrap sample, and set $\hat{\mathcal M}_{n,b}^*=\sqrt n\{\hat\Pi_{n,b}^*-\hat\Pi_n\}$. The number $B$ may be set to be $1000$ or larger if desired.
  \item Construct the rank estimator $\hat r_n$ for $\Pi_0$. For the two-step approach, choose $\beta\in(0,\alpha)$ (e.g., $\beta=\alpha/10$), sequentially test whether $\mathrm{rank}(\Pi_0)$ equals $0,1,...,k-1$ based on the test of \citet{Kleibergen_Paap2006rank} at the significance level $\beta$, and then let $\hat r_n=j^*$ if accepting $\mathrm{rank}(\Pi_0)=j^*$ is the first acceptance in the procedure, and $\hat r_n=k$ if no acceptance occurs. If $\hat r_n>r$, reject $\mathrm H_0$ and stop; otherwise, proceed to the next step. For the analytical approach, choose a small $\kappa_n$ (e.g., $\kappa_n=n^{-1/4}$), compute $\hat r_n$ according to \eqref{Eqn: rank estimator}, and directly proceed to the next step.
  \item Implement the singular value decomposition $\hat\Pi_n=\hat P_n\hat\Sigma_n\hat Q_n^\transpose$ and let $\hat P_{2,n}$ and $\hat Q_{2,n}$ be formed by the last $(m-\hat r_n)$ and $(k-\hat r_n)$ columns of $\hat P_n$ and $\hat Q_n$ respectively.
  \item Compute the smallest $k-r$ singular values of $\hat{P}_{2,n}^\transpose \hat{\mathcal M}_{n,b}^* \hat{Q}_{2,n}$ for each $b=1,\ldots,B$, and set the critical value to be $\lfloor B(1-\alpha+\beta)\rfloor$-th (for the two-step approach) or $\lfloor B(1-\alpha)\rfloor$-th (for the analytical approach) largest number in
\begin{align}\label{Eqn: critical value}
\sum_{j=r-\hat r_n+1}^{k-\hat{r}_{n}}\sigma^{2}_j(\hat{P}_{2,n}^\transpose \hat{\mathcal M}_{n,1}^* \hat{Q}_{2,n})~,\,\ldots~,\sum_{j=r-\hat r_n+1}^{k-\hat{r}_{n}}\sigma^{2}_j(\hat{P}_{2,n}^\transpose \hat{\mathcal M}_{n,B}^* \hat{Q}_{2,n})~.
\end{align}
Here, $\lfloor a\rfloor$ denotes the largest integer no larger than $a\in\mathbf R$.
\end{enumerate}

\noindent\underline{Step 3:} Reject $\mathrm H_0$ if and only if $n\phi(\hat\Pi_n)$ exceeds the critical value obtained in the previous step. Note that, for the two-step approach, this step may not be necessary if the rank estimator $\hat r_n$ in Step 2-(ii) is larger than $r$. We also stress that if the command reports the $p$-value of the two-step test in the second step, then the null hypothesis is rejected if and only if the $p$-value is smaller than $\alpha-\beta$ (as opposed to $\alpha$).

\section{Installation of the bootranktest package}\label{Sec: Installation}
The \texttt{bootranktest} command is available at the Statistical Software Components (SSC) archive. Our Stata package, \texttt{bootranktest}, can be installed from within Stata by typing \texttt{ssc install bootranktest}, which installs the command and the help files.

\section{The bootranktest command}\label{Sec: the command}
The syntax of \texttt{bootranktest} is as follows:

\noindent\texttt{bootranktest} (\textit{varlist1}) (\textit{varlist2}) [\textit{weight}] [if \textit{exp}] [in \textit{range}]\\
\indent[, \texttt{rank}(\#) \texttt{\underline{all}rank} \texttt{\underline{numb}oot}(\#) \texttt{\underline{b}eta}(\#) \texttt{\underline{k}appan}(\#) \\\indent \texttt{\underline{block}size}(\#) \texttt{partial}(\textit{varlist3}) \texttt{\underline{cl}uster}(\textit{varname}) \texttt{\underline{noc}onstant cfa}]

\subsection{Description}
\texttt{bootranktest} implements the tests described in Section \ref{Sec: Implementation} for the hypothesis in \eqref{Eqn: hypotheses1}, where $\Pi_0$ is the matrix of population coefficients in \eqref{Eqn: firststage} and $r$ is the hypothesized rank which may be specified in the option \texttt{rank}(\#) by user if necessary. To use this command, the user must specify two sets of variables in \eqref{Eqn: firststage}: \textit{varlist1} consists of variables in $Z$ and \textit{varlist2} consists of variables in $X$. A third set of variables, \textit{varlist3}, may be specified to include non-constant variables in $W$ as the intput for the option \texttt{partial}(\textit{varlist3}). The command allows \textit{varlist3} to be empty, in which case $W$ consists of only the constant unless the option \texttt{nonconstant} is specified. In the default case, $r=k-1$, $B=1000$, $\kappa_n = n^{-1/4}$, $\beta = 0.05/10$,  the wild bootstrap is implemented, and the result for the two-step approach is reported. The user may specify the \texttt{cfa} option to see the result for the analytical approach.

\subsection{Options}
\parindent=0pt
\parskip=\medskipamount

\hangindent=2em
\hangafter=1
\texttt{rank}(\#) specifies the hypothesized rank $r$, with default value $k-1$. Note that the value of $r$ must be strictly less than $k$.

\hangindent=2em
\hangafter=1
\texttt{\underline{all}rank} instructs the command to report all the results for $r=0,\ldots,k-1$.

\hangindent=2em
\hangafter=1
\texttt{\underline{numb}oot}(\#) specifies the number of bootstrap samples, with default value $1000$. In the default case, the command implements the wild bootstrap.

\hangindent=2em
\hangafter=1
\texttt{\underline{b}eta}(\#) specifies the value of $\beta$ that is needed to obtain $\hat{r}_{n}$ for the two-step approach, with default value $0.05/10$.

\hangindent=2em
\hangafter=1
\texttt{\underline{k}appan}(\#) specifies the value of $\kappa_n$ that is needed to obtain $\hat{r}_{n}$ for the analytical approach, with default value $n^{-1/4}$.

\hangindent=2em
\hangafter=1
\texttt{\underline{block}size}(\#) specifies the block size for block bootstrap that is implemented in the case of time series with serial dependence.

\hangindent=2em
\hangafter=1
\texttt{partial}(\textit{varlist3}) specifies the nonconstant variables in $W$.

\hangindent=2em
\hangafter=1
\texttt{\underline{cl}uster}(\textit{varname}) specifies the variable by which the observations are clustered. With this option specified, the command implements the cluster bootstrap.

\hangindent=2em
\hangafter=1
\texttt{\underline{noc}onstant} indicates that no constant variable is included in $W$.

\hangindent=2em
\hangafter=1
\texttt{cfa} instructs the command to report the result for the analytical approach, in addition to the result for the two-step approach.

\subsection{Stored results}
\texttt{bootranktest} stores the following in \texttt{r()}:

\noindent Scalars\\
\hangindent=2em
\hangafter=1
\begin{tabular}{lp{10cm}}
\texttt{r(cfa\_Teststat)} & Test statistic for the analytical approach\\
\texttt{r(cfa\_Pvalue)} & The p-value for the analytical approach\\
\texttt{r(cfa\_Rankestimate)} &$\hat{r}_{n}$ for the analytical approach\\
\texttt{r(cft\_Teststat)} & Test statistic in the second step of the two-step approach,\\
				& if rejection does not occur in the first step\\
\texttt{r(cft\_Pvalue)} & The p-value in the second step of the two-step approach,\\
				& if rejection does not occur in the first step\\
\texttt{r(cft\_Rankestimate)} &$\hat{r}_{n}$ for the two-step approach\\
\end{tabular}

\noindent Matrices\\
\hangindent=2em
\hangafter=1
\begin{tabular}{lp{10cm}}
\texttt{r(cfa\_rkmatrix)} & \hspace{0.65cm}Collection of the \texttt{\underline{all}rank} results for the analytical\\
				& \hspace{0.65cm}approach\\
\texttt{r(cft\_rkmatrix)} & \hspace{0.65cm}Collection of the \texttt{\underline{all}rank} results for the two-step\\
				& \hspace{0.65cm}approach\\
\end{tabular}

\noindent Macros\\
\hangindent=2em
\hangafter=1
\begin{tabular}{ll}
\texttt{r(cmd)} & \hspace{2.28cm}Improved test of matrix rank (name)\\
\end{tabular}

\section{Examples}\label{Sec: Examples}
We provide an example of tests for under-identification of the Klein consumption equation using the built-in Stata dataset \texttt{klein.dta} to demonstrate the basic functionality of the \texttt{bootranktest} command, and to compare to the Kleibergen-Paap test, which is implemented by the \texttt{ranktest} command \citep{Kleibergenetal_Ranktest_2020}. The Klein dataset consists of 22 time series observations (year 1920-1941), and out of this dataset, the variables of interests are consumption (\textit{consump}), private profits (\textit{profits}), total US wage bill (\textit{wagetot}), government spending (\textit{govt}), indirect bus taxes plus net export (\textit{taxnetx}), calendar year minus 1931 (\textit{year}), government wage bill (\textit{wagegovt}), lagged value of capital stock (\textit{capital1}), and total income/demand (\textit{totinc}).

\parindent=2em

\indent For reference, the Klein consumption equation is given as follows:
\begin{align}\label{KleinEqn}
\textit{consump}_t = \beta_0+ \beta_1 \cdot \textit{profits}_{t-1}+\beta_2 \cdot \textit{profits}_t +\beta_3 \cdot \textit{wagetot}_t + e_t,
\end{align}
for $t=2,\dots, 22$, in which the lagged $\textit{profits}_{t-1}$ is assumed to be the exogenous regressor, and $\textit{profits}_t$ and $\textit{wagetot}_t$ are assumed to be the endogenous regressors. The instruments for endogenous regressors are $\textit{govt}_t$, $\textit{taxnetx}_t$, $\textit{year}_t$, $\textit{wagegovt}_t$, $\textit{capital1}_t$, and $\textit{totinc}_{t-1}$, along with $\textit{profits}_{t-1}$ and the constant as the included exogenous regressors.

\indent Suppose that the goal is to test the relevance condition for the endogenous variables and IV's. Specifically, consider the first stage equation:
\begin{align}\label{KleinFirstStage}
\begin{bmatrix}
\textit{profits}_t \\
\textit{wagetot}_t
\end{bmatrix}
=
\Pi_0^{\transpose}
\begin{bmatrix}
\textit{govt}_t \\
\textit{taxnetx}_t \\
\textit{year}_t \\
\textit{wagegovt}_t \\
\textit{capital1}_t \\
\textit{totinc}_{t-1}
\end{bmatrix}
+
\Gamma_0^{\transpose}
\begin{bmatrix}
1 \\
\textit{profits}_{t-1}
\end{bmatrix}
+u_t,
\end{align}
and the null hypothesis is
\begin{align}\label{KleinHyp}
\mathrm H_0: \rank(\Pi_0) \leqslant 1.
\end{align}

\indent To import and prepare the data for for the use of time series operators, one uses the following commands:
\begin{stlog}
.   webuse klein, clear
{\smallskip}
.   tsset yr
        time variable:  yr, 1920 to 1941
                delta:  1 unit
{\smallskip}

\nullskip
\end{stlog}

The \texttt{bootranktest} results are:
\begin{stlog}
.   bootranktest (govt taxnet year wagegovt capital1 L.totinc) (profits wagetot) ///
>                 , partial(L.profits) cfa 
Test statistic in the second step of the two-step approach = 8.1005329
The p-value in the second step of the two-step approach = .031
(Note: the null hypothesis is rejected at alpha level if the p-value is smaller 
> than alpha-.005).
Test statistic for the analytical approach = 8.1005329
The p-value for the analytical approach = .632
{\smallskip}

\nullskip
\end{stlog}
The two-step approach yields a p-value that rejects  $H_0$ in \eqref{KleinHyp} at $5\%$ level, while the analytical approach yields a p-value that fail to reject the null at $5\%$ level

\indent Next, we change \eqref{KleinHyp} to
\begin{align}\label{KleinHyp2}
\mathrm H_0: \rank(\Pi_0) =0.
\end{align}
In addition, we assume first order lag, then the \texttt{bootranktest} results are
\begin{stlog}
.   bootranktest (govt taxnet year wagegovt capital1 L.totinc) (profits wagetot) ///
>                 , partial(L.profits) block(2) cfa rank(0)
Test statistic in the second step of the two-step approach = 69.488582
The p-value in the second step of the two-step approach = .63
(Note: the null hypothesis is rejected at alpha level if the p-value is smaller 
> than alpha-.005).
Test statistic for the analytical approach = 69.488582
The p-value for the analytical approach = .63
{\smallskip}

\nullskip
\end{stlog}
which are autocorrelation robust when \texttt{blocksize()} is specified with value equals to order of lag plus one to account for serial dependence.

\indent To compare, we report the outputs for the \texttt{ranktest} command below in the order of testing the null hypotheses \eqref{KleinHyp} and \eqref{KleinHyp2}, respectively:
\begin{stlog}
.   ranktest (profits wagetot) (govt taxnet year wagegovt capital1 L.totinc) ///
>                 , partial(L.profits) robust
{\smallskip}
Kleibergen-Paap rk LM test of rank of matrix
  Test statistic robust to heteroskedasticity
Test of rank=  0  rk=   18.07  Chi-sq( 12) pvalue=0.113693
Test of rank=  1  rk=    4.92  Chi-sq(  5) pvalue=0.425234
{\smallskip}

\nullskip
\end{stlog}

\begin{stlog}
.   ranktest (profits wagetot) (govt taxnet year wagegovt capital1 L.totinc) ///
>                 , partial(L.profits) bw(2) robust null
{\smallskip}
Kleibergen-Paap rk LM test of rank of matrix
  Test statistic robust to heteroskedasticity and autocorrelation
  Kernel: Bartlett   Bandwidth: 2
Test of rank=  0  rk=    9.88  Chi-sq( 12) pvalue=0.626575
{\smallskip}
\nullskip
\end{stlog}

\bibliographystyle{sj}
\bibliography{mybibliography}

\begin{aboutauthors}
Qihui Chen is an Assistant Professor at the Chinese University of Hong Kong, Shenzhen.

Zheng Fang is an Assistant Professor at Texas A\&M University.

Xun Huang is a graduate student at San Francisco State University.
\end{aboutauthors}

\clearpage
\end{document}